\documentclass[twocolumn,english,showabstract, showkeys, showpacs]{revtex4}
\usepackage[T1]{fontenc}
\usepackage[latin9]{inputenc}
\usepackage{color}
\usepackage{array}
\usepackage{textcomp}
\usepackage{amstext}
\usepackage{graphicx}
\usepackage{esint}

\makeatletter

\DeclareRobustCommand{\greektext}{%
  \fontencoding{LGR}\selectfont\def\encodingdefault{LGR}}
\DeclareRobustCommand{\textgreek}[1]{\leavevmode{\greektext #1}}
\DeclareFontEncoding{LGR}{}{}

\providecommand{\tabularnewline}{\\}

\@ifundefined{textcolor}{}
{%
 \definecolor{BLACK}{gray}{0}
 \definecolor{WHITE}{gray}{1}
 \definecolor{RED}{rgb}{1,0,0}
 \definecolor{GREEN}{rgb}{0,1,0}
 \definecolor{BLUE}{rgb}{0,0,1}
 \definecolor{CYAN}{cmyk}{1,0,0,0}
 \definecolor{MAGENTA}{cmyk}{0,1,0,0}
 \definecolor{YELLOW}{cmyk}{0,0,1,0}
 }

\makeatother

\usepackage{babel}

\begin{document}

\title{Magnetic Behavior in RRh$_{3}$X (\textcolor{black}{R = rare earths;}\textcolor{blue}{{}
}\ X=B, C) Compounds}

\author{Devang A. Joshi}
\email{devang@tifr.res.in}

\author{Neeraj Kumar, A. Thamizhavel
and S. K. Dhar}

\address{{\large Department of Condensed Matter Physics and Materials Science,
Tata Institute of Fundamental Research, }\textcolor{black}{\large Colaba,
}{\large Mumbai 400 005, India.}}

\begin{abstract}
We report on the magnetic behavior of RRh$_{3}$B (R = La, Ce, Pr,
Nd, Gd, Tb and Tm) and RRh$_{3}$C (R = La, Ce, Pr and Gd) compounds
crystallizing in the cubic perovskite type structure with space group
Pm3m. The heat capacity data on Pauli-paramagnetic LaRh$_{3}$B and
LaRh$_{3}$C indicate a high frequency vibrating motion of boron and
carbon atoms in the unit cell. Ce is in $\alpha$-like nonmagnetic
state in both the compounds. Pr compounds show a dominant crystal
field effect with a nonmagnetic singlet ground state in PrRh$_{3}$B
and a nonmagnetic quadrupolar doublet in PrRh$_{3}$C. Compounds with
other rare earths order ferromagnetically at low temperatures except
TmRh$_{3}$B in which the zero field evolution of magnetic interactions
is relatively more complicated. The electrical resistivity of GdRh$_{3}$B
decreases with increasing temperature in the paramagnetic state in
the vicinity of T$_{\mathrm{C}}$, which is rarely seen in ferromagnets.
The behavior is discussed to be arising due to the short range spin
fluctuation and a possible contribution from Fermi surface geometry.
\end{abstract}



\maketitle

\section{INTRODUCTION}

The existence of a large number of borides and carbides of rare earth
compounds forming in the anti-perovskite structure is well known \cite{Holleck,Rogl1}.
Boron and carbon occupy the body-centered (\textonehalf{}, \textonehalf{},
\textonehalf{}) position of the FCC unit cell in these compounds.
Dhar et al. had reported the synthesis of RPd$_{3}$B$_{\mathrm{x}}$
(0 \ensuremath{\le} x <1) by alloying boron with RPd$_{3}$ (R = rare
earth) compounds which have the AuCu$_{3}$-type FCC structure \cite{Dhar1}.
The introduction of smaller atoms like boron and carbon leads in some
cases to interesting changes in the physical properties of the parent
compounds. For example, boron expands the unit cell and induces a
change of valence from intermediate valent in CePd$_{3}$ to trivalent
in CePd$_{3}$B$_{\mathrm{x}}$ (x > 0.25) and from trivalent in EuPd$_{3}$
to mixed valent in EuPd$_{3}$B$_{\mathrm{x}}$ (x $\sim$ 0.5) \cite{Dhar2,Dhar3}.
Evidence for charge ordering in EuPd$_{3}$B was inferred from the
variation of $^{151}$Eu Mössbauer spectra with temperature \cite{Dhar4}.
The synthesis of perovskite type RRh$_{3}$B$\mathrm{_{x}}$ has also
been reported in the literature \cite{Dhar5,Takei}, though in this
case only CeRh$_{3}$ is known to exist in AuCu$_{3}$-type structure.
The Curie-Weiss, paramagnetic behavior of the susceptibility of RRh$_{3}$B
compounds between 300 and 77 K and $^{151}$Eu Mössbauer measurements
in EuRh$_{3}$B indicating trivalent behavior of Eu ions have been
reported \cite{Dhar5}. Rogl and De-Long investigated some ternary
borides for superconductivity but found HfRh$_{3}$B$_{1-\mathrm{x}}$
and YRh$_{3}$B$_{1-\mathrm{x}}$ alloys non-superconducting down
to 1.5 K \cite{Rogl2}. A narrowing of the 5\textit{\textcolor{black}{f}}
\ band due to the \textcolor{black}{hybridization} between U 5\textit{f
}electrons and Rh \textit{d} electrons was inferred in the perovskite
URh$_{3}$B from heat capacity and susceptibility, further supported
by band structure calculations \cite{Dunlap}. More recently, cubic
perovskite MgNi$_{3}$C was found to become superconducting at T $\sim$
8 K \cite{He}. In the recent past, RRh$_{3}$B$_{\mathrm{x}}$ compounds
have been studied primarily with regard to the variation of hardness
with boron content \cite{Shishido,Shishido1}. In this communication,
we report the magnetic properties of RRh$_{3}$B (R = La, Ce, Pr,
Nd, Gd, Tb and Tm), based on \textcolor{black}{the}\textcolor{blue}{{}
}\ magnetization, heat capacity and electrical resistivity data measured
down to 1.7 K. For comparison, RRh$_{3}$C compounds for R = La, Ce,
Pr and Gd were also synthesized and their magnetic behavior examined.

\section{EXPERIMENTAL}

The compounds were prepared by melting the stoichiometric amount\textcolor{black}{s}
of constituent elements in an arc furnace on a water cooled cooper
hearth under an inert atmosphere of argon. The purity of the starting
material is R (99.9~wt.~\%), Rh (99.9~wt.~\%), B (99.5~wt.~\%)
and C (99.9999~wt.~\%). The compounds were annealed at 800 \textdegree{}C
for a week and checked with x-ray diffraction using Cu K\textgreek{a}
radiation. Magnetization measurements were carried out using a \textcolor{black}{Quantum
Design MPMS-5 superconducting quantum interference device magnetometer
and Oxford Instruments vibrating sample magnetometer} while the electrical
resistivity was recorded on an automated home-built set up. The heat
capacity was measured using a \textcolor{black}{physical property
measurement system (PPMS, Quantum Design). }

\section{RESULTS AND DISCUSSION}

The RRh$_{3}$B and RRh$_{3}$C series of compounds form in a \textcolor{black}{cubic
perovskite} structure with space group Pm3m \cite{Holleck}. The series
of compounds form for almost all the rare earths in contrast to the
parent RRh$_{3}$, which forms only in case of Ce.\textcolor{black}{{}
\ Hence an additional atom (boron or carbon) is required to stabilize
the perovskite structure. A minimum amount of boron is needed to stabilize
the RRh$_{3}$B$_{x}$ phase, the magnitude of x depends upon the
rare earth R \cite{Shishido1}. We believe a similar situation should
also hold for alloying with carbon.} \textcolor{black}{In order to
confirm the phase homogeneity of }%
\begin{figure}
\textcolor{black}{\includegraphics[width=0.5\textwidth]{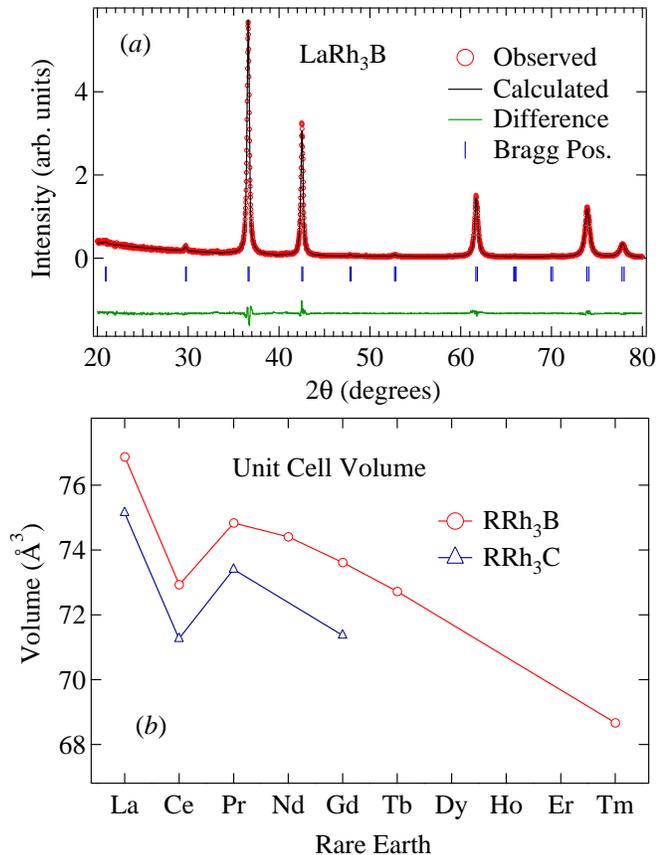}}

\textcolor{black}{\caption{(Color online)\label{Fig. Rietveld-Unit-Cell}\textcolor{blue}{{} }\textcolor{black}{(a)
Powder x-ray diffraction pattern of LaRh$_{3}$B at room temperature.
The solid line through the experimental data points is the Rietveld
refinement profile calculated for the cubic LaRh$_{3}$B. The Bragg
position and the difference of the observed and calculated profile
is also shown. (b) Unit cell volume of RRh$_{3}$B and RRh$_{3}$C
compounds plotted against their corresponding rare-earth constituents,
the line joining the symbols is guide to the eye. }}
}
\end{figure}
\textcolor{black}{{} RRh$_{3}$B and RRh$_{3}$C compounds prepared
in this work,} with proper lattice and crystallographic parameters,
a Rietveld analysis of the observed x-ray patterns of all the compounds
was done using the Fullprof program. The refined plot of LaRh$_{3}$B
is shown in Fig. 1a. The obtained lattice parameters are presented
in \textcolor{black}{Ta}ble 1 and the corresponding unit cell volume\textcolor{black}{s
are} plotted in Fig. 1b. The unit cell volume decreases as we move
to higher rare earths in accordance with the lanthanide contraction.
For CeRh$_{3}$B and CeRh$_{3}$C, however, the value is below that
expected on the basis of lanthanide contraction, indicating an $\alpha$-like
nonmagnetic state of Ce ions in these compounds.%
\begin{table}
\begin{tabular}{ccccc}
 &  &  &  & \tabularnewline
\hline
\hline 
Compound & Lattice Parameters &  &  & \tabularnewline
 & \textit{a }($\textrm{\AA}$) & T$_{\mathrm{C}}$ (K) & $\mathrm{\mu}_{eff}$($\mathrm{\mu}_{\mathrm{B}}$) & $\mathrm{\theta_{p}}$ (K)\tabularnewline
\hline
LaRh$_{3}$B & 4.252 & P-P & - & -\tabularnewline
CeRh$_{3}$B & 4.178 & P-P & - & -\tabularnewline
PrRh$_{3}$B & 4.214 & P & 3.58 & -7\tabularnewline
NdRh$_{3}$B & 4.206 & Below 2 & 3.62 & 3\tabularnewline
GdRh$_{3}$B & 4.191 & 12 & 7.88 & 12\tabularnewline
TbRh$_{3}$B & 4.174 & 6 & 9.5 & 5\tabularnewline
TmRh$_{3}$B & 4.095 & 4 & 7.57 & -1\tabularnewline
LaRh$_{3}$C & 4.22 & P-P & - & -\tabularnewline
CeRh$_{3}$C & 4.146 & P-P & - & -\tabularnewline
PrRh$_{3}$C & 4.187 & P & 3.58 & -11\tabularnewline
GdRh$_{3}$C & 4.148 & 3.5 & 7.9 & 3\tabularnewline
\hline
\hline 
 &  &  &  & \tabularnewline
\end{tabular}

\caption{Lattice parameters, \textcolor{black}{magnetic}\textcolor{blue}{{} }\ ordering
temperature, effective moment and paramagnetic Curie temperature for
RRh$_{3}$B and RRh$_{3}$C compounds. P-P: Pauli-paramagnetic and
P: Paramagnetic.}

\end{table}
 The unit cell volume of RRh$_{3}$B is larger than that of the corresponding
RRh$_{3}$C compounds, and it may be due to the larger metallic radius
of the boron atom (0.88 $\textrm{\AA}$) compared to that of the carbon
atom (0.77 $\textrm{\AA}$).

\subsection{LaRh$_{3}$B and LaRh$_{3}$C}

\begin{figure}
\includegraphics[width=0.5\textwidth]{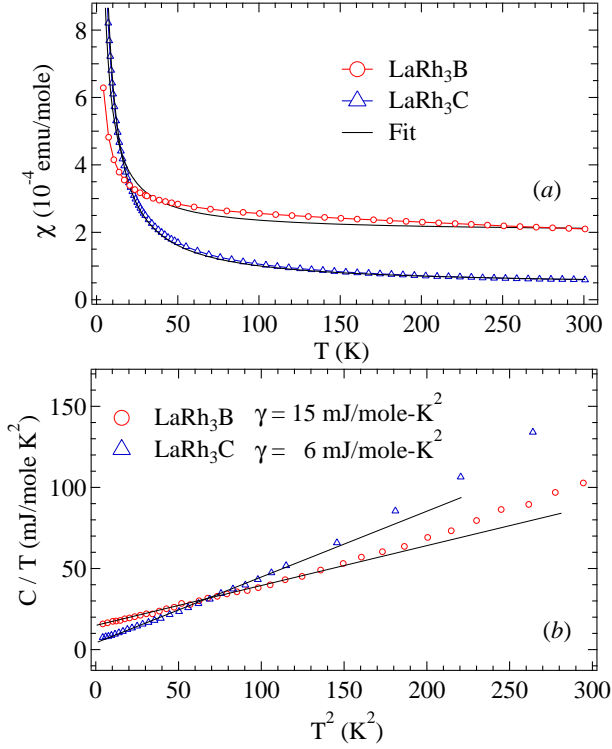}

\caption{(Color online)\textcolor{black}{\label{Fig. Susceptibility_La} (a)
Magnetic susceptibility of LaRh$_{3}$B and LaRh$_{3}$C; the solid
lines represent a fit to the modified Curie-Weiss law. The line joining
the data points is guide to eye. (b) C/T vs T$^{2}$ curve for LaRh$_{3}$B
and LaRh$_{3}$C compounds. The solid lines represent a fit to an
equation described in the text.}}

\end{figure}
We first discuss the behavior of nonmagnetic\textcolor{blue}{, }\textcolor{black}{reference
compounds }LaRh$_{3}$B and LaRh$_{3}$C. The magnetic susceptibilities
of LaRh$_{3}$B and LaRh$_{3}$C are shown in Fig. 2a. Both the compounds
exhibit a Pauli-paramagnetic behavior. There is an increase in the
susceptibilities at low temperatures, which may be due to the paramagnetic
impurity within the samples\textcolor{black}{,}\textcolor{blue}{{} }\textcolor{black}{arising
from the presence of paramagnetic ions in the starting materials.}
In order to estimate the temperature independent susceptibility ($\chi_{0}$)
the modified Curie-Weiss law $(\chi=\chi_{0}+N\mu_{eff}^{2}/3k_{B}(T-\theta_{P}))$
\textcolor{black}{was fitted to the data, the various parameters have
t}heir usual meaning. The obtained values of $\chi_{0}$ are $2.0\times10^{-4}$
and $0.4\times10^{-4}$ emu/mole with effective moment of 0.17 and
0.22 $\mu_{B}$ respectively for LaRh$_{3}$B and LaRh$_{3}$C. Such
a low value of effective moment indicates that the upturn at low temperature
is due to the presence of \textcolor{black}{paramagnetic} impurity\textcolor{black}{{}
\ ions}. The value of $\chi_{0}$ for LaRh$_{3}$C is much less than
that of LaRh$_{3}$B. Considering the fact that the Pauli-paramagnetism
arises from the conduction electron polarization, it is possible that
replacing boron by carbon \textcolor{black}{shifts the Fermi level
to a region of low density of states. This is supported by the low
temperature heat ca}pacity \textcolor{black}{data plotted as C/}T
vs T$^{2}$ in Fig. 2b. The \textcolor{black}{magnitude of the e}lectronic
contribution $\gamma$\textcolor{black}{,}\textcolor{blue}{{} }\textcolor{black}{obtained
by fitting the expression}\textcolor{blue}{{} \ ${\color{black}C/T=\gamma+\beta T^{2}}$
}\textcolor{black}{to the data,} is 15 and 6 mJ/mole K$^{2}$ for
LaRh$_{3}$B and LaRh$_{3}$C respectively. Since $\gamma$ is proportional
to the density of states at the Fermi level, a reduction in the value
of $\gamma$ for LaRh$_{3}$C indicate\textcolor{black}{s a} decrease
in the density of states $N\left(E_{F}\right)$ at the Fermi level.\textcolor{black}{{}
}\textcolor{blue}{\ }\textcolor{black}{The magnitude \ of }$N\left(E_{F}\right)$\textcolor{black}{{}
\ can be obtained using the free electron relation}

\textcolor{black}{\begin{equation}
\gamma=\frac{2}{3}\:\pi^{2}\: k_{B}^{2}\: N(E_{F})\label{Eq: Density}\end{equation}
where $k_{B}$ is the Boltzmann constant. Substituting the value of
$\gamma$} for LaRh$_{3}$B and LaRh$_{3}$C\textcolor{black}{, the
density of states at the Fermi level is found to be 43.4 Ry$^{-1}$atom$^{-1}$
and 17.4 Ry$^{-1}$atom$^{-1}$ respectively.}

\begin{figure}
\includegraphics[width=0.5\textwidth]{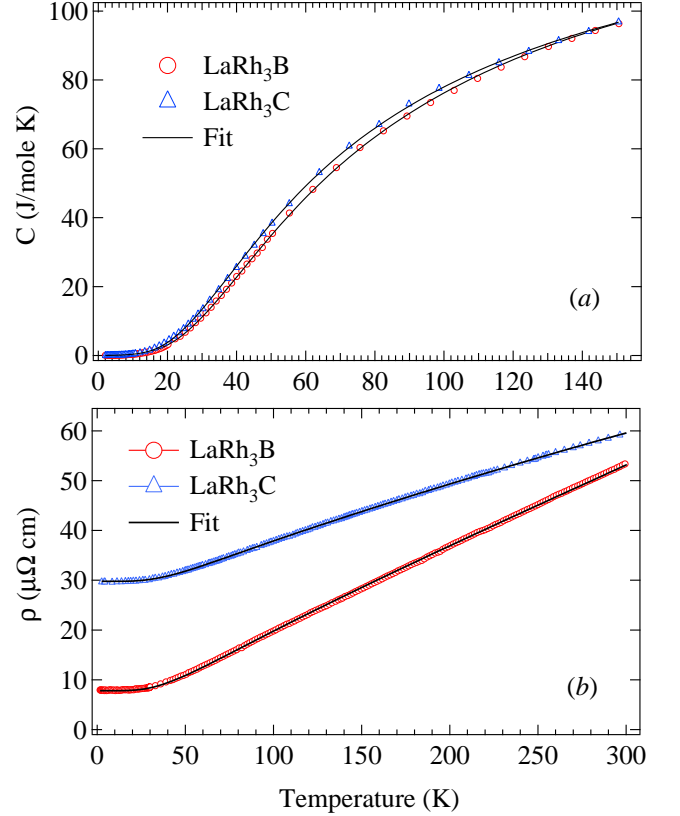}

\caption{(Color online) \label{Fig. Resistivity}\textcolor{black}{(a) Heat
capacity as a function of temperature for LaRh$_{3}$B and LaRh$_{3}$C
with a fit to a combined Einstein and Debye contribution. (b) Resistivity
behavior of LaRh$_{3}$B and LaRh$_{3}$C with a fit to Bloch-Gr$\mathrm{\ddot{u}}$neisen
relation.}}

\end{figure}
The heat capacity of LaRh$_{3}$B and LaRh$_{3}$C \textcolor{black}{between
1.7 and 150 K is plotted} in Fig. 3a. The heat capacity of LaRh$_{3}$C
is seen to be higher than that of LaRh$_{3}$B above 10 K but at lower
temperatures the heat \textcolor{black}{capacity of the latter dominates
because of its higher electronic contribution. The thermal variation
of the heat capacity of these two compounds could not be described
by either the Debye or Einstein model separately, but could be well
described by combined Einstein and Debye contributions as shown by
the solid lines in Fig. 3a. The total heat capacity in such case is
given by}\begin{equation}
C_{Tot}=\gamma T+(C_{E}+C_{D})\label{Eq. Total-HC}\end{equation}
\textcolor{black}{where the first term represents \ the electronic
contribution, The second term includes Einstein contribution C$_{E}$
and Debye contribution C$_{D}$. The Einstein contribution is given
by \begin{equation}
C_{E}=\sum_{n'}3n_{En'}R\frac{y^{2}e^{y}}{\left(e^{y}-1\right)^{2}}\label{Eq. Einstein}\end{equation}
where }$y$= $\Theta_{En'}/T$, $\Theta_{E}$ is the Einstein temperature,
$n'$ is the summation over the different Einstein temperatures, $R$
is the gas constant and $n_{E}$ is the number of Einstein oscillators.\textcolor{black}{{}
\ The Debye contribution is given by}

\textcolor{black}{\begin{equation}
C_{D}=9n_{D}R\left(\frac{T}{\Theta_{D}}\right)^{3}\intop_{0}^{\Theta_{D}/T}\frac{x^{4}e^{x}dx}{\left(e^{x}-1\right)^{2}}\label{Eq. Debye}\end{equation}
where} $x$= $\Theta_{D}/T$. $\Theta_{D}$ is the Debye temperature
and $n_{D}$ is the number of Debye oscillators. Iterative fit to
the Eq. 2 was performed by using the values of electronic contribution
$\gamma$ as estimated above and fixing the number of atoms $n_{D}$
and $n_{E}$ for a particular fit, allowing both $\Theta_{En'}$ and
$\Theta_{D}$ to vary as fitting parameters. A good fit to the heat
capacity of LaRh$_{3}$B and LaRh$_{3}$C over the entire range of
temperature was obtained by assigning three Debye characteristic atoms
($n_{D}$ = 3) with $\Theta_{D}=\mathrm{287}$ K plus two Einstein
characteristic atoms ($n_{E1}=n_{E2}=1$) with $\Theta_{E1}=150$
K and $\Theta_{E2}=569$ K for LaRh$_{3}$B and $\Theta_{D}=\mathrm{272}$
K, $\Theta_{E1}=135$ K and $\Theta_{E2}=569$ K for LaRh$_{3}$C.
The total number of atoms ($n_{D}+n_{E}=5$) accounts for the five
atoms of each LaRh$_{3}$B and LaRh$_{3}$C compound. The description
of heat capacity in terms of a combination of acoustic and optical
contribution can readily be understood by assigning the Rh atoms in
the unit cell to three Debye characteristic modes and the remaining
to the Einstein modes. In the Einstein mode of vibration, $\Theta_{E1}$
<\textcompwordmark{}< $\Theta_{E2}$, since La is much larger in size
(\textcolor{black}{about 100 \%)} compared to B or C \textcolor{black}{atoms,
thus it can be expected to vibrate with a lower natural frequency
compared to that of the latter ones. H}ence $\Theta_{E1}$ corresponds
to the vibration of La atoms and $\Theta_{E2}$ to that of B or C
atoms. The change in the Debye temperature $\Theta_{D}$ and Einstein
temperature $\Theta_{E1}$ from LaRh$_{3}$B to LaRh$_{3}$C can be
understood to be due to the change in the lattice parameters, whereas
a constant value of $\Theta_{E2}$ for both the compounds could not
be understood but can tentatively be assigned to similar bonding strength
of the boron and carbon with the neighboring atoms. 

The resistivity of LaRh$_{3}$B \textcolor{black}{exhibits a metallic
like behavior down to 30 K (Fig. 3b), leveling off at low temperature
with a residual resistivity of 7.8 $\mu\Omega\mathrm{\, cm}$. The
resistivity of LaRh$_{3}$C also shows a similar behavior but with
a small negative curvature at low temperature (between 75 and 150
K) and a residual resistivity of 30 $\mu\Omega\,\mathrm{cm}$. A possible
source of negative curvature in the resistivity of nonmagnetic materials
is due to the scattering of conduction electrons from }\textit{\textcolor{black}{s\ }}\textcolor{black}{{}
to }\textit{\textcolor{black}{d}}\textcolor{black}{{} \ band \cite{Mott}.
The scattering is proportional to the density of states in the }\textit{\textcolor{black}{d}}\textcolor{black}{{}
\ band. Thus the substitution of C in place of B results in the shifting
of the Fermi surface in the vicinity of }\textit{\textcolor{black}{d}}\textcolor{black}{{}
\ band. Also the resistivity of LaRh$_{3}$C is higher than that
of LaRh$_{3}$B due to the decrease in the density of states at the
Fermi level. }The resistivity of both the compoun\textcolor{black}{ds
c}ould be described by the modified Bloch-Gr$\mathrm{\ddot{u}}$neisen
relation given by \begin{equation}
\rho(T)=\rho_{0}+4\Theta_{D}R\left(\frac{T}{\Theta_{D}}\right)^{5}\intop_{0}^{\Theta_{D}/T}\frac{x^{5}dx}{\left(e^{x}-1\right)(1-e^{-x})}-KT^{3}\label{Eq: Bloch-Gru}\end{equation}
where $x$= $\Theta_{D}/T$, $\Theta_{D}$ is the Debye temperature,
$\rho_{0}$ is the temperature independent residual resistivity and
$R\,\mathrm{and}\, K$ are the coefficients of the phonon contribution
to the resistivity (second term) and the Mott \textit{s}-\textit{d}
inter band scattering (third term) respectively. The fit to the resistivity
curve\textcolor{black}{s} yields $\Theta_{D}=223$ K, $\rho_{0}=7.8\,\mu\Omega\,\mathrm{cm}$,
$R=0.156\,\mu\Omega\,\mathrm{cm}$~K$^{-1}$ and $K=0\,\mu\Omega\mathrm{\, cm\,}$K$^{-3}$
for LaRh$_{3}$B and $\Theta_{D}=226$~K, $\rho_{0}=30\,\mu\Omega\mathrm{\, cm}$,
$R=0.11\,\mu\Omega\,\mathrm{cm}$~K$^{-1}$ and $K=0.46\,10^{-7}\mu\Omega\,\mathrm{cm}$~K$^{-3}$
for LaRh$_{3}$C. The finite value of the coefficient $K$ for LaRh$_{3}$C
supports the explanation of the small curvature in the resistivity
curve\textcolor{black}{{} \ proposed above. }The Debye temperature
in the Bloch-Gr$\mathrm{\ddot{u}}$neisen relation often differs from
the value of $\Theta_{D}$ obtained from heat capacity data \textcolor{black}{because}\textcolor{blue}{{}
}the former takes into account only the longitudinal phonons \cite{Gopal}.

\subsection{CeRh$_{3}$B and CeRh$_{3}$C}

\begin{figure}
\includegraphics[width=0.5\textwidth]{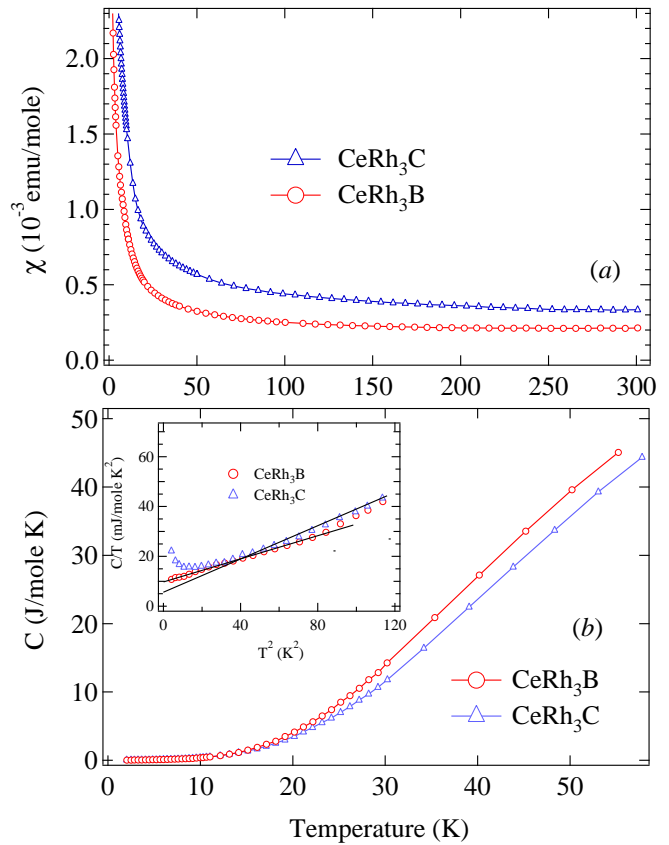}

\caption{(Color online) \textcolor{black}{\label{Fig. Ce_Mag}(a) Magnetic
susceptibility of CeRh$_{3}$B and CeRh$_{3}$C. The line joining
the data points are guide to eye. (b) Heat capacity of CeRh$_{3}$B
and CeRh$_{3}$C compounds with the inset showing C/T vs T$^{2}$
curve for both compounds. The solid lines in the inset represent a
fit as described in text.}}

\end{figure}
\textcolor{black}{From a comparison of their lattice constants with
neighboring La and Pr analogs, CeRh$_{3}$B and CeRh$_{3}$C are most
likely intermediate valance compounds. This is well corroborated by
their magnetic susceptibility, as shown in Fig. 4a. The susceptibility
of CeRh$_{3}$B at 300 K is $\sim$} $2.3\times10^{-4}$ emu/mole
and remains nearly temperature independent down to about 50 K, consistent
with Pauli paramagnetic behavior. It is seen that the susceptibility
of CeRh$_{3}$B is comparable to that of LaRh$_{3}$B and consistent
with the non-magnetic nature of a nearly tetravalent or $\alpha$-like
state of Ce ions. The upturn in susceptibility at low temperatures
is attributed to the presence of trace paramagnetic ions present in
the starting constituents. A similar behavior is also observed in
CeRh$_{3}$C with $\chi\sim$ $3.5\times10^{-4}$ emu/mole at 300
K, \textcolor{black}{comparable to that of CeRh$_{3}$B.} Even though
the density of states at the Fermi level for CeRh$_{3}$B is higher
than that of CeRh$_{3}$C (inferred from $\gamma$ value as described
below), the higher value of the susceptibility of CeRh$_{3}$C can
\textcolor{black}{be attributed to the slightly different 4}\textit{\textcolor{black}{f}}\textcolor{black}{{}
\ occupation in these two compounds. }

The heat capacity of the compound\textcolor{black}{s} is shown in
Fig. 4b. The heat capacity of CeRh$_{3}$B is less than that of CeRh$_{3}$C
below $\sim$ 11 K\textcolor{black}{{} \ but exceeds the latter at
higher temperatures. T}he behavior is in sharp contrast to that of
the corresponding La compounds. The reason may be due to the chang\textcolor{black}{es
i}n the Debye and Einstein characteristic temperatur\textcolor{black}{es
of these two sets of compounds. The electronic contribution to the
heat capacity estimated from the linear part of the C/T vs T$^{2}$
curve for CeRh$_{3}$B and CeRh$_{3}$C below 10 K is $\sim$10 and
5 mJ/mole K$^{2}$. The higher value for the boride is in agreement
with the results on the}\textcolor{blue}{{} }\ corresponding La compounds.

\subsection{PrRh$_{3}$B and PrRh$_{3}$C}

\begin{figure}
\includegraphics[width=0.5\textwidth]{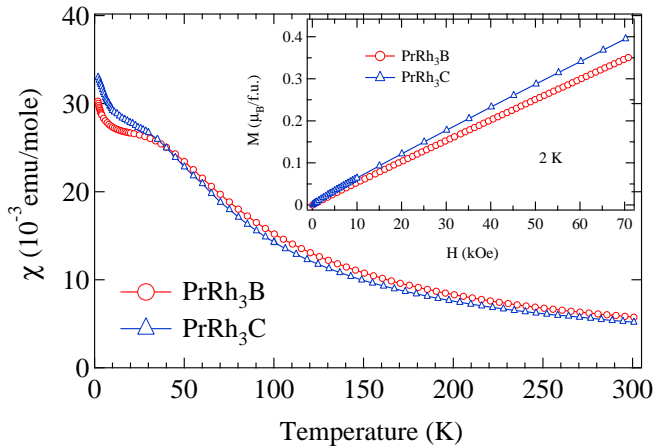}

\caption{(Color online) \label{Fig. MT-MH_Pr}\textcolor{black}{Magnetic susceptibility
of PrRh$_{3}$B and PrRh$_{3}$C, with the inset showing their magnetic
isotherm at 2 K. The lines joining the data points are guide to eye.}}

\end{figure}
The susceptibility of PrRh$_{3}$B and PrRh$_{3}$C is shown in Fig.
\ref{Fig. MT-MH_Pr}. The susceptibility of both the compounds increases
with decreasing temperature, followed by a nearly temperature-independent
behavior at $\approx$ 25 K, followed by a further increase at low\textcolor{black}{er
temperatures.} The magnetic isotherm of the compoun\textcolor{black}{ds
}at 2 K (inset of Fig. 5a) shows a linear behavior up to 70 kOe\textcolor{black}{,}
\textcolor{black}{attaining a moment of $\approx$ 0.35 and 0.4 $\mu_{B}$/f.u.
for PrRh$_{3}$B and PrRh$_{3}$C, respectively. The behavior of the
susceptibility and a relatively low moment at 2 K in high fields indicates
a paramagnetic state of these compounds with a strong crystal electric
field effect. }Such a behavior is \textcolor{black}{not un}common
in the case of Pr compounds. The crystal field can split the J = 4
ground state of the Pr$^{3+}$ ion, resulting in a nonmagnetic singlet\cite{Buschow}
or a \textcolor{black}{nonmagnetic quadrupolar doublet ground state
(PrPb$_{3}$, PrMg$_{3}$ and PrPtBi: \cite{Bucher,Tanida,Suzuki}),
if the excha}nge interactions do not exceed a certain critical value
\cite{Bleaney}. Such nonmagnetic ground states would give rise to
a temperature-independent, Van-Vleck susceptibility at low temperatures.
Hence the nearly temperature-independent behavior of the susceptibility
(Fig. 5a) at T $\approx$ 25 K for both the compounds arises from
a nonmagnetic ground state. The upturn in the susceptibility at low
temperatures is \textcolor{black}{due to some trace paramagnetic impurities
in the compounds. The fitting of the Curie-Weiss law to the inverse}
susceptibility furnish\textcolor{black}{es an }effective moment of
3.58 $\mu_{B}$ for both compounds, equal to the theoretically expected
value for the Pr$^{3+}$ ion. The paramagnetic Curie temperature\textcolor{black}{s}
are -7 and -11 K for PrRh$_{3}$B and PrRh$_{3}$C, respectively.
\textcolor{black}{The negative values are normally taken to imply
an antiferromagnetic type of interaction, but since there is no magnetic
transition down to 1.8 K they must result from crystal electric field
effect. This is in contrast with the ferromagnetic interaction present
in rest of the compounds with magnetic rare earth ions R.}

\textcolor{black}{}%
\begin{figure}
\textcolor{black}{\includegraphics[width=0.5\textwidth]{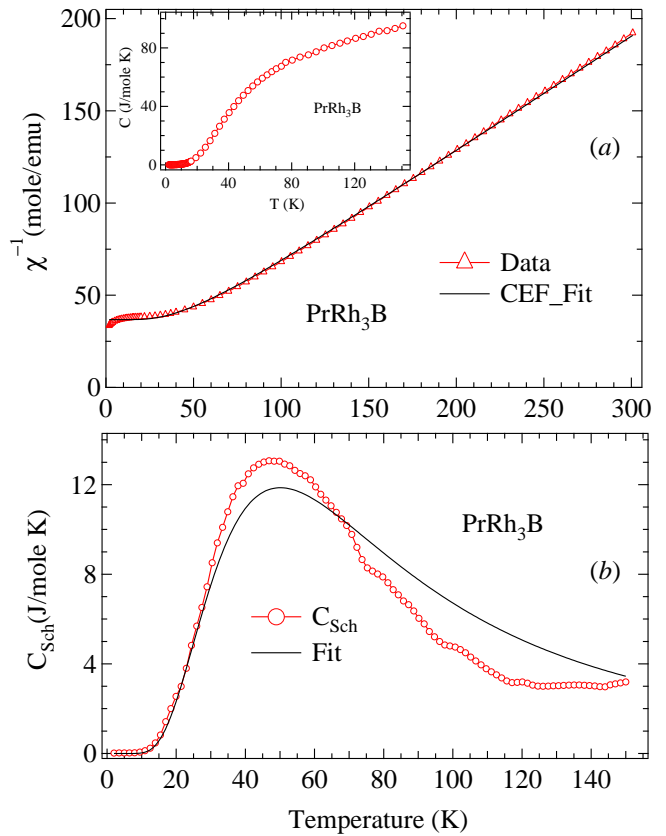}}

\textcolor{black}{\caption{(Color online)\textcolor{black}{{} \label{Fig. PrB_CEF_Sch}(a): Inverse
susceptibility as a function of temperature for PrRh$_{3}$B. The
solid line through the data points represent a fit to CEF model as
described in text. The inset shows the heat capacity of PrRh$_{3}$B.
(b): Schottky contribution to the heat capacity of PrRh$_{3}$B with
a solid line representing the theoretically simulated Schottky curve.}}
}

\end{figure}
\textcolor{black}{The heat capacities of PrRh$_{3}$B and PrRh$_{3}$C
are shown as insets to Fig. \ref{Fig. PrB_CEF_Sch}a and Fig. \ref{Fig. PrC_CEF_Sch}a,
respectively. The behavior is similar to that of a \ nonmagnetic
compound. The Schottky contribution to the heat capacity of both the
compounds was isolated by subtracting the heat capacity of the La
analogue, after taking into account the slight difference in the atomic
masses of La and Pr, and is \ shown in Fig. \ref{Fig. PrB_CEF_Sch}b
and Fig. \ref{Fig. PrC_CEF_Sch}b, respectively. Both the compounds
show a Schottky peak in their heat capacity arising from the thermal
variation of the fractional occupation of the crystal electric field
levels. }

\textcolor{black}{A set of CEF levels which reproduce the Schottky
heat capacity reasonably well may plausibly be taken to represent
the actual CEF splitting in these two compounds. Towards this goal
the susceptibility of the two compounds was analyzed on the basis
of a CEF model including the exchange parameter. Crystal field analysis
was done taking into account the cubic symmetry of the Pr$^{3+}$
ion in PrRh$_{3}$B and PrRh$_{3}$C. The CEF Hamiltonian for a cubic
point group symmetry is given by \cite{Hutchings}\begin{equation}
H_{CEF}^{cub.}=B_{4}^{0}\left(O_{4}^{0}+5O_{4}^{4}\right)+B_{6}^{0}\left(O_{6}^{0}-21O_{6}^{4}\right)\end{equation}
where $B_{l}^{m}$ and $O_{l}^{m}$ are the CEF parameters and the
Stevens operators, respectively. The diagonalization of the above
Hamiltonian was done using the computer simulation to fit the inverse
susceptibility of the compounds, as shown in Fig. \ref{Fig. PrB_CEF_Sch}a
and \ref{Fig. PrC_CEF_Sch}a respectively}%
\begin{figure}
\includegraphics[width=0.5\textwidth]{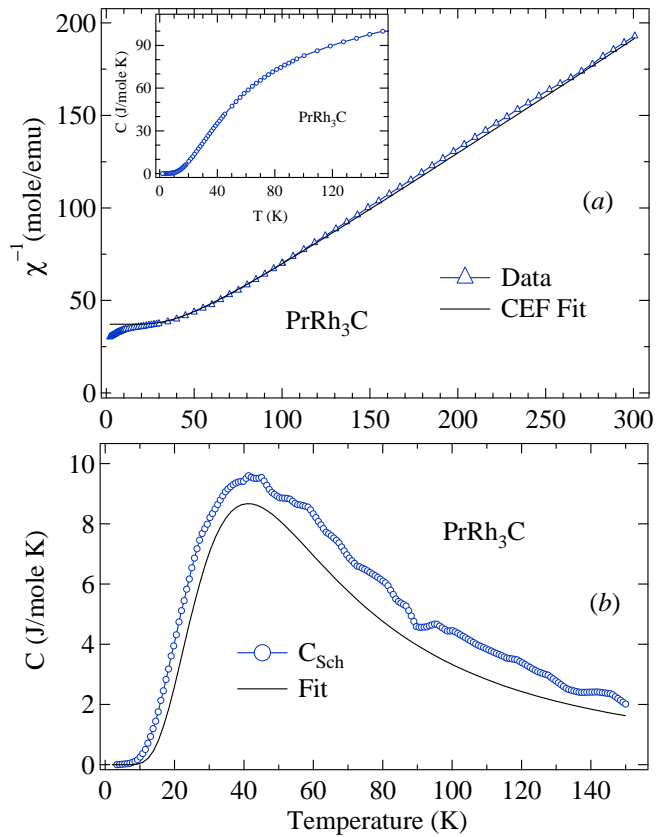}

\textcolor{black}{\caption{(Color online) \label{Fig. PrC_CEF_Sch}(a): \textcolor{black}{Inverse
susceptibility as a function of temperature for PrRh$_{3}$C. The
solid line represent a fit to CEF model as described in text. The
inset shows the heat capacity behavior of PrRh$_{3}$C. (b): Schottky
contribution to the heat capacity of PrRh$_{3}$C with solid line
representing the simulated Schottky curve.}}
}

\end{figure}
\textcolor{black}{. The CEF parameters obtained from the fit are $B_{4}^{0}$
= -0.079 K and $B_{6}^{0}$ = 0.0005 K for PrRh$_{3}$B and $B_{4}^{0}$
= 0.0121 K and $B_{6}^{0}$ = -0.00135 K for PrRh$_{3}$C. The susceptibility
could be fitted to a set of values of the crystal electric field parameters,
but only those parameters are taken into account which could also
fit the experimentally obtained Schottky anomaly as well. The crystal
field split energy levels obtained from CEF parameters are shown in
Fig.8. The ground state for PrRh$_{3}$B was found to be a singlet
whereas that of PrRh$_{3}$C is a doublet. Since the carbide is not
ordering magnetically at least down to 1.8 K, the ground state is
most likely a quadrupolar nonmagnetic doublet. These \ energy levels
were \ used to calculate the Schottky contribution from \ the equation\begin{equation}
C_{Sch}\left(T\right)=R\left[\frac{\sum_{i}g_{i}e^{-E_{i}/T}\sum_{i}g_{i}E_{i}^{2}e^{-E_{i}/T}-\left[\sum_{i}g_{i}E_{i}e^{-E_{i}/T}\right]^{2}}{T^{2}\left[\sum_{i}g_{i}e^{-E_{i}/T}\right]^{2}}\right]\end{equation}
where R is the gas constant, E$_{\mathrm{i}}$ is the CEF energy level
in units of temperature and g$_{\mathrm{i}}$ the corresponding degeneracy.
The curves obtained are shown in Fig. \ref{Fig. PrB_CEF_Sch}b and
Fig. \ref{Fig. PrC_CEF_Sch}b, respectively for PrRh$_{3}$B and PrRh$_{3}$C.
There is only a qualitative agreement between the observed and the
calculated Schottky contribution. The discrepancy may arise due to
the uncertainty in isolating the pure phonon contribution as the phonon
spectra of PrRh$_{3}$B and PrRh$_{3}$C may not be identical to that
of LaRh$_{3}$B and LaRh$_{3}$C, respectively. The peak positions,
however, nearly match indicating that our inferred CEF level scheme
is fairly close to the actual splitting in both compounds scheme.}%
\begin{figure}
\includegraphics[width=0.5\textwidth]{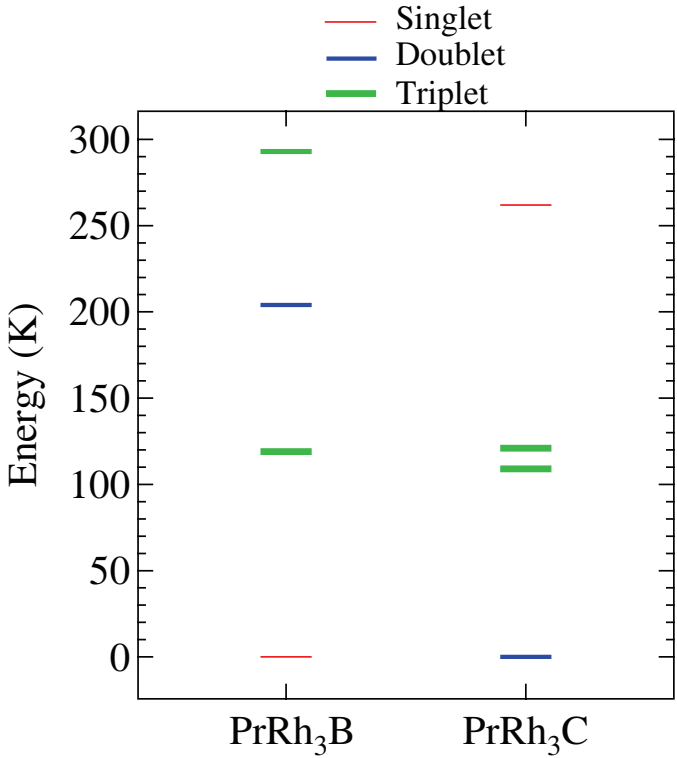}

\textcolor{black}{\caption{(Color online)\textcolor{black}{{} Crystal field split energy levels
of PrRh$_{3}$B and PrRh$_{3}$C estimated from a CEF fit to inverse
susceptibility.}}
}

\end{figure}
\textcolor{black}{{} }

\subsection{GdRh$_{3}$B and GdRh$_{3}$C}

\begin{figure}
\includegraphics[width=0.5\textwidth]{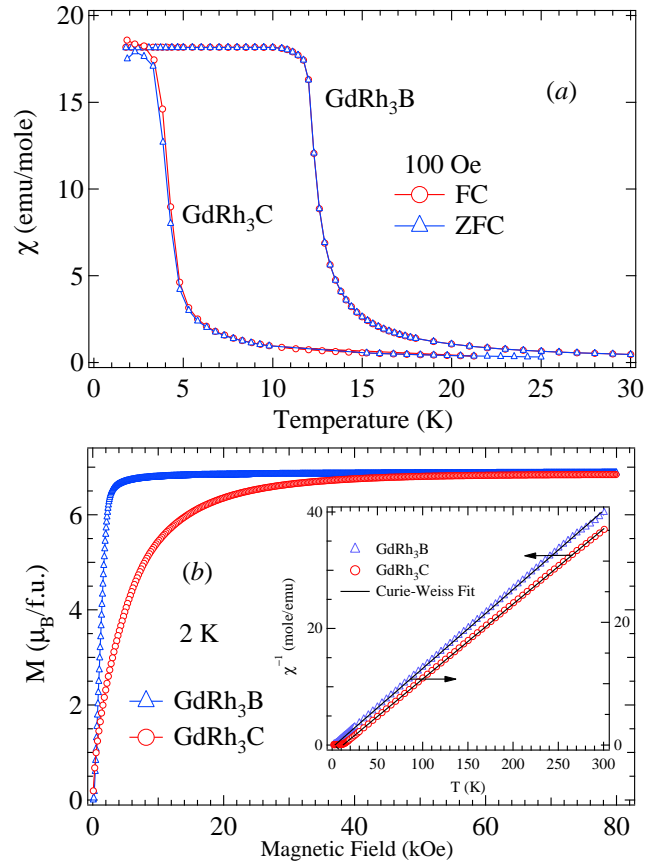}

\caption{(Color online) \label{Fig. Gd_Mag}a: \textcolor{black}{Magnetic susceptibility
of GdRh$_{3}$B and GdRh$_{3}$C under zero field cooled and field
cooled conditions. b: Magnetic isotherm of both the compounds at 2
K. The inset shows the inverse magnetic susceptibility of both the
compounds with solid lines representing a Curie-Weiss fit.}}

\end{figure}
The results on GdRh$_{3}$B and GdRh$_{3}$C are presented next as
Gd is an \textit{S}-state ion and the CEF effects are negligible in
the first order approximation. The magnetic susceptibility of both
the compounds under zero field cooled (ZFC) and field cooled (FC)
conditions at 100 Oe is shown in Fig. \ref{Fig. Gd_Mag}a. GdRh$_{3}$B
and GdRh$_{3}$C order ferromagnetically at 12 and\textcolor{black}{{}
\ 3.5 K respectively. Substituting carbon in place of boron decreases
the ordering temperature substantially. Since the Gd$^{3+}$ ion is
unaffected by the crystal field, the lower magnetic ordering temperature
in the carbide is likely due to a decrease in the density of states
at the Fermi level, inferred from the magnitude of $\gamma$ in La
and Ce compounds when boron is replaced by carbon. In case of GdRh$_{3}$B,
the susceptibilities under FC and ZFC conditions coincide with each
other, whereas a slight difference is observed for GdRh$_{3}$C. The
behavior of the magnetic isotherm of both the compounds at 2 K (Fig.
\ref{Fig. Gd_Mag}b) is as expected for a ferromagnetically ordered
compound. The magnetization for GdRh$_{3}$B increases sharply at
low fields and saturates quickly, compared to t}hat of GdRh$_{3}$C
which saturates above 40 kOe. The saturation moments obtained for
GdRh$_{3}$B and GdRh$_{3}$C respectively are 6.87 and 6.84 $\mu_{B}/\mathrm{f.u.}$
at 2 K and 80 kOe, close to the theoretically expected value of the
Gd$^{3+}$ ion ( 7 $\mu_{B}$).

\begin{figure}
\includegraphics[width=0.5\textwidth]{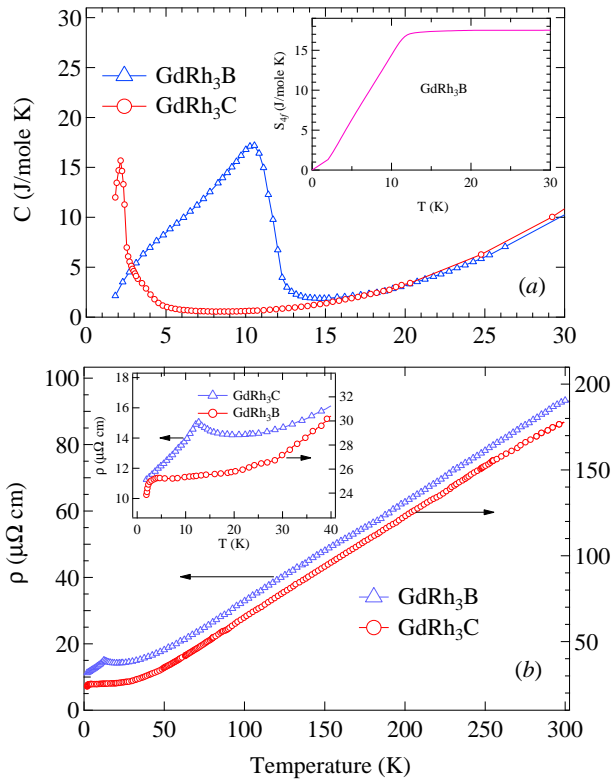}

\caption{(Color online) \label{Fig._HC_Rho_Gd}\textcolor{black}{a: Low temperature
heat capacity of GdRh$_{3}$B and GdRh$_{3}$C with the inset showing
the calculated entropy for GdRh$_{3}$B. The line joining the data
points are guide to eye. b: Resistivity behavior of GdRh$_{3}$B and
GdRh$_{3}$C. The inset shows the low temperature part for both compounds.
The arrows in both plots point to their respective scales.}}

\end{figure}
 The heat capacity of both the compounds shows an anomaly at the magnetic
ordering temperature. (Fig. \ref{Fig._HC_Rho_Gd}a). In GdRh$_{3}$C
the heat capacity \textcolor{black}{shows a broad shoulder between
2 and 5 K followed by a sharp peak at 2.2 K in contrast to a relatively
sharp upturn in GdRh$_{3}$B. The heat capacity data in GdRh$_{3}$C
suggest a relatively complicated evolution of the magnetic interactions
leading to full magnetic ordering at 2.2 K, which may be responsible
for the slight difference in the ZFC and FC plots and for the relatively
slow saturation of the magnetization in GdRh$_{3}$C at 2 K. As regards
the latter point it may also be noted that the isothermal magnetization
plot at 2 K is taken close to the magnetic ordering temperature. The
magnetic contribution to the heat capacity of GdRh$_{3}$B was isolated
using the procedure described above and the entropy at 30 K is 17.6
J/mole K, close to the expected value of $Rln(2J+1)$.}

\begin{figure}
\includegraphics[width=0.5\textwidth]{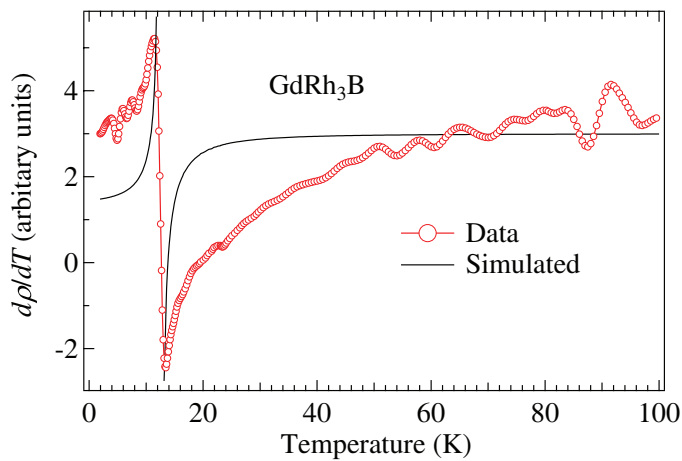}

\caption{(Color online) \label{Fig. first-derivative}\textcolor{black}{The
first derivative of resistivity for GdRh$_{3}$B indicating a discontinuity
at the ordering temperature and the solid line represents the theoretically
simulated curve to support the behavior. }}

\end{figure}
The resistivity of both the compounds is shown in Fig. \ref{Fig._HC_Rho_Gd}b.
with the inset showing the low temperature part. The resistivity at
high temperature shows a metallic behavior for both the compounds.
For GdRh$_{3}$B the resistivity decreases from $\approx$ 90 $\mu\Omega$
cm at room temperature to $\approx$ 10 $\mu\Omega$ cm at 2 K in
contrast to a significant drop in the resistivity of GdRh$_{3}$C
from $\approx$ 180 $\mu\Omega$ cm at room temperature to $\approx$
24 $\mu\Omega$ cm at 2 K.\textcolor{black}{{} The higher resistivity
of GdRh$_{3}$C compared to that of GdRh$_{3}$B supports the low
density of states at the Fermi level in the former. At low temperature
(inset of Fig. \ref{Fig._HC_Rho_Gd}b) the resistivity of GdRh$_{3}$C
shows a drop at the ordering temperature of the compound due to the
gradual disappearance of the spin disorder resistivity. Interestingly
the resistivity of GdRh$_{3}$B increases below 20 K up to the ordering
temperature in the paramagnetic regime showing a cusp at the ordering
temperature. Such a behavior of resistivity is common in antiferromagnets
but very rarely seen in ferromagnets ( eg., }\textit{\textcolor{black}{c}}\textcolor{black}{{}
direction of Gd metal, La(Fe, Si)$_{13}$ and Fe$_{3}$Pt \cite{Maezawa,Palstra,Viard}
). In the case of antiferromagnets it has been argued that as one
approaches T$_{\mathrm{C}}$ from above, the growth of spin correlations
lead to large-wave-vector q fluctuations (q: magnetic reciprocal lattice
vector) which increase the resistivity due to spin fluctuations as
the temperature is decreased \cite{Balberg}. In ferromagnets, such
behavior arises for various reasons, in case of \ La(FeSi)$_{13}$
and Fe$_{3}$Pt the behavior arises due to the lattice softening associated
with the Invar effect \cite{Weiss}. In case of Gd metal, the presence
of small \ Fermi surface caliper in the }\textit{\textcolor{black}{c}}\textcolor{black}{{}
\ direction is responsible for the experimentally observed cusp in
the resistivity along the }\textit{\textcolor{black}{c}}\textcolor{black}{\ -axis
\cite{Geldart,Freeman}. In addition to these effects, a theory by
Kim \cite{Kim} shows that an anomaly in the resistivity of the ferromagnetic
metals or alloys at their ordering temperature arises due to the scattering
of conduction electrons by \ short range spin fluctuations. \ Kawatra
}\textit{\textcolor{black}{et. al.}}\textcolor{black}{{} \ \cite{Kawatra}
using Kims theory \cite{Kim} showed that presence of short range
spin fluctuations cause a sharp discontinuity in the $d\rho/dT$ curve
at the ordering temperature of the compound. The $d\rho/dT$ curve
for GdRh$_{3}$B is shown in Fig. 11. It shows a sharp anomaly and
a change in sign at the ordering temperature. Also the sharp anomaly
in $d\rho/dT$ curve is well described by the the simulated curve
(Fig. 11) using equations described by Kim \cite{Kim}. This shows
that short range spin fluctuations is the major cause for the cusp
at the ordering temperature. Here we do not rule out the contribution
from Fermi surface geometry, but a proper estimation of the Fermi
surface and single crystal data are required to confirm.}

\subsection{NdRh$_{3}$B, TbRh$_{3}$B and TmRh$_{3}$B}

\begin{figure}
\includegraphics[width=0.5\textwidth]{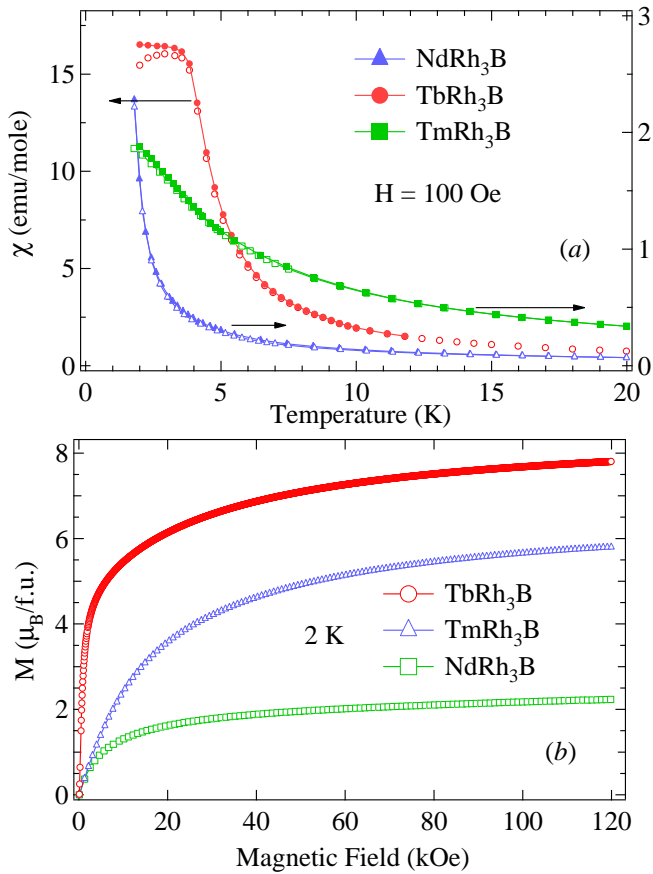}\caption{(Color online) \label{Fig_Mag_NdTbTm}a: Magnetic susceptibility of
NdRh$_{3}$B, TbRh$_{3}$B and TmRh$_{3}$B under ZFC (unfilled symbols)
and FC (Filled symbols) conditions. The arrows indicates their respective
axis. b: Magnetic isotherms of the same compounds at 2 K. }

\end{figure}
\textcolor{black}{{} T}he magnetic behavior of NdRh$_{3}$B, TbRh$_{3}$B
and TmRh$_{3}$B \textcolor{black}{is described next.} \textcolor{black}{The
selection of a Tm compound from the series is because of its expected
behavior to form nonmagnetic singlet ground state.} In the paramagnetic
state the fit of the Curie-Weiss law (not shown) to the susceptibility
of NdRh$_{3}$B and TbRh$_{3}$B furnishes $\mu_{eff}$= 3.62 and
9.5 $\mu_{B}$ and paramagnetic Curie temperature $\theta_{P}$= 3
and 5 K\textcolor{black}{,}\textcolor{blue}{{} }\textcolor{black}{respectively.}
The effective moments are close to their respective free ion moment,
and the positive value of $\theta_{P}$ indicates the ferromagnetic
type of interaction between the moments. \textcolor{black}{The ferromagnetic
nature of the magnetic ordering in TbRh$_{3}$B below 6 K and \ below
2 K in NdRh$_{3}$B is apparent from the low-field magnetization plots
shown in Fig. \ref{Fig_Mag_NdTbTm}a. A slight deviation between the
ZFC and FC plots below 4 K in TbRh$_{3}$B \ and the field dependence
of the isothermal magnetization at 2 K (Fig. \ref{Fig_Mag_NdTbTm}b)\ 
suggests the presence of magnetic anisotropy arising from the CEF
effects. On the other hand both the FC and ZFC curves of NdRh$_{3}$B
in the paramagnetic state coincide with each other as expected. The
magnetic isotherm for TbRh$_{3}$B (Fig. \ref{Fig_Mag_NdTbTm}b) is
typical of a ferromagnetically ordered compound, with a moment of
7.8 $\mu_{B}/\mathrm{f.u.}$ at 2 K and 120 kOe, close to the saturation
moment of 9 $\mu_{B}/\mathrm{{\color{blue}{\color{black}Tb}}}$ .
Si}milar behavior \textcolor{black}{is seen in NdRh$_{3}$B at 2 K,
with a magnetic moment of $\approx$ 2.3 $\mu_{B}/\mathrm{f.u.}$
at 120 kOe. }%
\begin{figure}
\textcolor{black}{\includegraphics[width=0.5\textwidth]{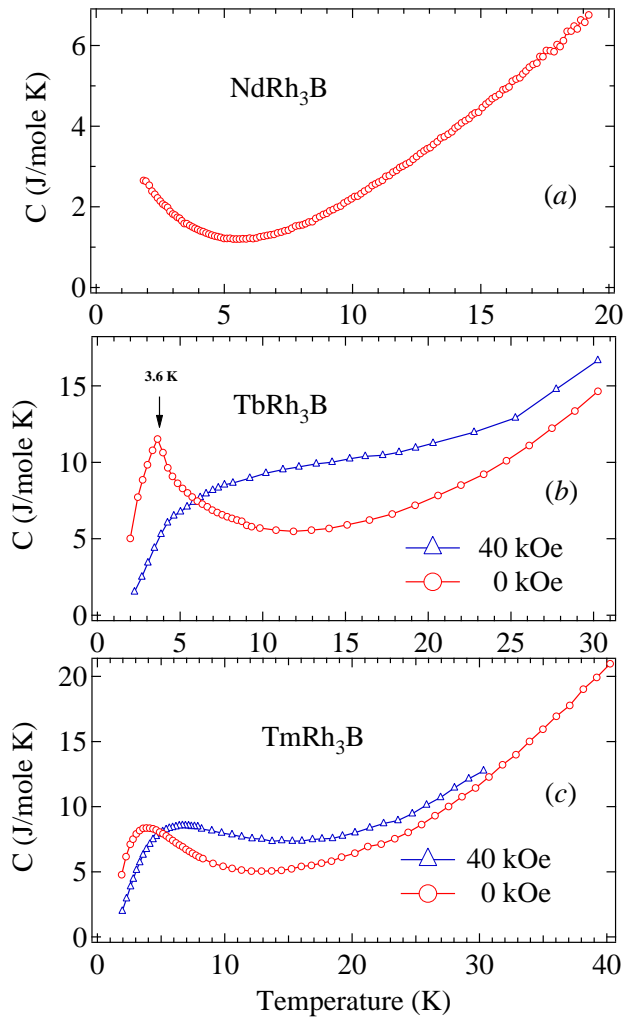}}

\textcolor{black}{\caption{(Color online) \label{Fig. HC_NdTbTm}\textcolor{black}{Heat capacity
data of NdRh$_{3}$B, TbRh$_{3}$B and TmRh$_{3}$B. Heat capacity
of TbRh$_{3}$B and TmRh$_{3}$B in a field of 40 kOe are also shown.
The line joining the data points are guide to eye.}}
}
\end{figure}
\textcolor{black}{{} The electrical resistivity of TbRh$_{3}$B (Fig.
\ref{Fig Res_NdTbTm}) shows a clear anomaly at 6 K, decreasing sharply
at lower temperatures. However, the heat capacity shows a sharp peak
at 3.6 K (Fig. \ref{Fig. HC_NdTbTm}b). In the paramagnetic state
the heat capacity in zero field exhibits a relatively long tail suggestive
of extended short range order in the paramagnetic state. Therefore,
the anomaly at 6 K in the resistivity of TbRh$_{3}$B may signify
the onset of strong short range order than truly long range magnetic
ordering. The ferromagnetic nature of the magnetic correlations is
further suggested by the effect of magnetic field (40 kOe) on the
heat capacity of TbRh$_{3}$B (Fig. \ref{Fig. HC_NdTbTm}b). The peak
at 3.6 K vanishes and the entropy due to the ordering is shifted to
higher temperatures. The heat capacity of NdRh$_{3}$B (Fig. \ref{Fig. HC_NdTbTm}a)
shows an upturn below 5 K attaining a value of nearly 3 J/mole K at
the lowest temperature of 1.7 K. The upturn is a precursor to the
magnetic ordering at lower temperature. }The resistivity of NdRh$_{3}$B
shows a metallic behavior down to 2 K (Fig. \ref{Fig Res_NdTbTm})
with no signature of low temperature magnetic ordering.

\begin{figure}
\includegraphics[width=0.5\textwidth]{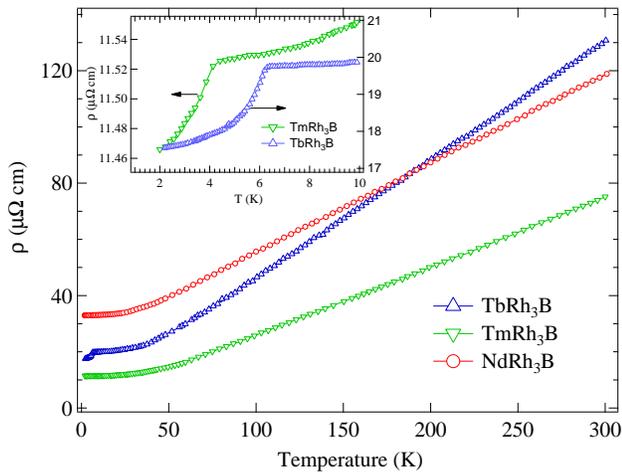}

\caption{(Color online) \label{Fig Res_NdTbTm}\textcolor{black}{Resistivity
behavior of NdRh$_{3}$B, TbRh$_{3}$B and TmRh$_{3}$B. The inset
shows the low temperature part for TbRh$_{3}$B and TmRh$_{3}$B.
The lines connecting the data points is guide to the eye. }}

\end{figure}
\textcolor{black}{The susceptibilities of TmRh$_{3}$B under ZFC and
FC modes (Fig. \ref{Fig_Mag_NdTbTm}a) nearly coincide with each other.
Unlike the Gd, Tb and Nd analogs there is no sharp rise but a change
in the slope at $\approx$ 4 K. The magnetic isotherm of the compound
at 2 K (Fig. \ref{Fig_Mag_NdTbTm}a) has a behavior typical of ferromagnets,
with a magnetic moment of $\approx$ 5.8 $\mu_{B}/\mathrm{f.u.}$
at 120 kOe. The Curie-Weiss fit to the inverse susceptibility in the
paramagnetic region yields $\mu_{eff}$= 7.6 $\mu_{B}$ and $\theta_{P}$=
-1 K. The heat capacity and resistivity data are shown in Fig. \ref{Fig. HC_NdTbTm}c
and \ref{Fig Res_NdTbTm}. The heat capacity shows a broad peak centered
at $\approx$ 4 K with a long tail in the paramagnetic region similar
to that of TbRh$_{3}$B. The effect of the applied magnetic field
on the heat capacity of TmRh$_{3}$B is similar to that in TbRh$_{3}$B,
the broad peak getting shifted to higher temperature by a few degrees.
The resistivity too shows a drop at $\approx$ 4 K but the drop in
the resistivity is not so significant, it just drops by just 0.08
$\mu\Omega\mathrm{\,\, cm}$ between 4 and 2 K. The low-field susceptibility
together with magnetic heat capacity and resistivity measurements
indicate the presence of a complicated magnetic structure.}

\section{Conclusion}

In conclusion, we have studied the magnetic behavior of RRh$_{3}$B
(R = La, Ce, Pr, Nd, Gd, Tb and Tm) and RRh$_{3}$C (R = La, Ce, Pr
and Gd) compounds. The compounds form in perovskite type cubic structure
with space group Pm3m. LaRh$_{3}$B and LaRh$_{3}$C show Pauli-paramagnetic
behavior. Substitution of C in place of B causes decrease in the density
of states at the Fermi level. Ce is in $\alpha$-like state in both
the compounds. Pr compounds shows a dominant crystal field effect
with a nonmagnetic singlet ground state in PrRh$_{3}$B and a nonmagnetic
quadrupolar doublet in PrRh$_{3}$C. Compounds with other rare earths
order ferromagnetically at low temperatures. Resistivity of GdRh$_{3}$B
shows the presence of short range spin fluctuations, which is rarely
seen in ferromagnets. TbRh$_{3}$B and TmRh$_{3}$B order ferromagnetically
at 6 and 4 K respectively. NdRh$_{3}$B has an ordering temperature
below 2 K.

\end{document}